\documentclass[aps,prapplied,twocolumn,letterpaper,superscriptaddress,10pt]{revtex4-2}
\usepackage{amssymb,amsthm,amsmath,amsfonts}
\usepackage{graphicx,ulem,enumerate,mathptmx}
\usepackage[pdftex,dvipsnames,usenames]{xcolor}
\usepackage[colorlinks=true,urlcolor=blue,citecolor=blue,linkcolor=blue]{hyperref}

\begin{document}

\title{Tailored frequency conversion makes infrared light visible for streak cameras}

\author{Carolin L\"uders}
    \email{carolin.lueders@tu-dortmund.de}
	\affiliation{Experimentelle Physik 2, Technische Universit\"at Dortmund, D-44221 Dortmund, Germany}
	
\author{Jano Gil-Lopez}
    \email{jangil@campus.uni-paderborn.de}
	\affiliation{Integrated Quantum Optics, Institute for Photonic Quantum Systems (PhoQS), Universit\"at Paderborn, Warburger Str. 100, 33098 Paderborn, Germany}
	
\author{Markus Allgaier}
    \email{markusa@uoregon.edu}
    \affiliation{Integrated Quantum Optics, Institute for Photonic Quantum Systems (PhoQS), Universit\"at Paderborn, Warburger Str. 100, 33098 Paderborn, Germany}
	\affiliation{Department of Physics and Oregon Center for Optical, Molecular,and Quantum Science, University of Oregon, Eugene, Oregon 97403, USA}
	
\author{Benjamin Brecht}
    \email{benjamin.brecht@uni-paderborn.de}
	\affiliation{Integrated Quantum Optics, Institute for Photonic Quantum Systems (PhoQS), Universit\"at Paderborn, Warburger Str. 100, 33098 Paderborn, Germany}

\author{Marc A\ss{}mann}
    \email{marc.assmann@tu-dortmund.de}
	\affiliation{Experimentelle Physik 2, Technische Universit\"at Dortmund, D-44221 Dortmund, Germany}
	
\author{Christine Silberhorn}
    \email{christine.silberhorn@uni-paderborn.de}
	\affiliation{Integrated Quantum Optics, Institute for Photonic Quantum Systems (PhoQS), Universit\"at Paderborn, Warburger Str. 100, 33098 Paderborn, Germany}
	
\author{Manfred Bayer}
    \email{manfred.bayer@tu-dortmund.de}
	\affiliation{Experimentelle Physik 2, Technische Universit\"at Dortmund, D-44221 Dortmund, Germany}

\begin{abstract}
	Streak cameras are one of the most common and convenient devices to measure pulsed emission e.g. from semiconductor lightsources with picosecond time resolution. However, they are most sensitive in the visible range and possess low or negligible efficiency in the infrared and telecom regime. In this work, we present a frequency conversion based on sum-frequency generation that converts infrared to visible signals while preserving their temporal properties, making them detectable with a streak camera. We demonstrate and verify the functionality of our device by converting the emission from a quantum dot laser. 
\end{abstract}

\date{\today}
\maketitle

%%%%%%%%%%%%%%%%%%%%%%%%%%%%%%%%%%%%%%%%%%%%%%%%%%%%
%%%%%%%%%%%%%%%%%%%%%%%%%%%%%%%%%%%%%%%%%%%%%%%%%%%%
%%%%%%%%%%%%%%%%%%%%%%%%%%%%%%%%%%%%%%%%%%%%%%%%%%%%
\section{Introduction}
Many applications require measuring the temporal properties of optical emission from semiconductors: Verifying single photon emission  \cite{Somashi2016} or quantifying quantum coherence \cite{Lueders2021} for quantum information tasks; analysing carrier and phonon dynamics \cite{Othonos1998}; observing the lasing dynamics of new laser candidates \cite{Zhang2020} or studying cooperative effects in quantum dots such as superradiance \cite{Jahnke2016} and superfluorescence \cite{Nasu2020}.\newline
For these tasks, detection methods with a temporal resolution in the range of pico- to femtoseconds are needed. These can be distinguished into techniques that measure the pulse form via intensity autocorrelation (e.g., optical autocorrelation \cite{Armstrong1967}, spectral shearing interferometry (SPIDER) \cite{Anderson2008}, frequency-resolved optical gating (FROG) \cite{Kane1993}; these methods typically require large field intensities), techniques working at the single photon level but not delivering short pulse forms (e.g. the Hanbury-Brown Twiss setup for $g^{(2)}$ measurements \cite{Brown1956}) and techniques resolving (ultra-)short pulse forms with single photon sensitivity (optical Kerr gating \cite{Gilabert1987}, frequency up-conversion sampling \cite{Mahr1975}, electro-optic shearing interferometry \cite{Davis2018}, and streak cameras \cite{Campillo1983}). Also there are various types of single-photon detectors with time resolutions on the order of 100~ps \cite{Hadfield2009}. Another kind of information is obtained by pump-probe schemes measuring time-resolved reflectivity or absorption of the material \cite{Othonos1998}. \newline 
While a plethora of techniques allow access to the temporal envelope of pulses at the single-photon level, only streak cameras \cite{Assmann2009,Wiersig2009} allow simultaneous access to other degrees of freedom, e.g. spatial \cite{Gao2014} or spectral in combination with a spectrometer. Especially in the semiconductor community, streak cameras are popular, as they allow for measuring photon correlations with ps time resolution, resolving the small coherence times of semiconductor light sources \cite{Takemura2012,Schmutzler2014,Adiyatullin2015,Adiyatullin2017}. \newline   
However, measuring technologically relevant wavelengths beyond 1000~nm (e.g., the telecommunication wavelengths around 1310~nm and 1550~nm, or mid-infrared wavelengths beyond 2 microns) poses additional challenges for these methods. The wavelength range of Kerr gating is limited to around 1100~nm by material properties and the wavelength of the gating pulse \cite{Schmidt2003, Appavoo2014}; streak cameras are constrained by the photocathode material and do not possess sufficient quantum efficiency in the telecom range \cite{Allgaier2018} (e.g. conventional S1 photocathodes have a radiant sensitivity of ca $2\cdot10^{-4}$~mA/W at 1500~nm, corresponding to a quantum efficiency of $1.6\cdot10^{-7}$ \cite{Hamamatsu2018}, whereas InP/InGaAs photocathodes have a radiant sensitivity of ca 1~mA/W at 1500~nm, corresponding to a quantum efficiency of $8\cdot10^{-4}$ \cite{Hamamatsu2015}); and electro-optic shearing interferometry relies on single-photon detection. Of all types of single-photon detectors, superconducting nanowire single-photon detectors are most promising, combining sensitivity for telecom wavelength together with timing jitters on the order of 20~ps \cite{Hadfield2009,Zadeh2021}. \newline
Apart from these, for the infrared, techniques based on frequency up-conversion sampling are the only viable candidates to date. In frequency up-conversion sampling, the investigated light is mixed with a bright pulsed pump laser in a nonlinear optical material and converted to the sum-frequency of the involved fields. The intensity of the up-converted light can be measured with any detector while varying the time delay between the signal field and pump pulses, thereby reconstructing the temporal shape of the signal with a time resolution given by the pump pulse duration. In semiconductor physics, this method has been applied e.g. to carrier dynamics in carbon nanotubes \cite{Ma2004}, superfluorescent emission of excitons in ZnTe \cite{Dai2011} and to $g^{(2)}$ measurements of a confined polariton structure \cite{Delteil2019}. \newline
While frequency up-conversion can be applied to a wide range of wavelengths, given a suitable nonlinear material, pump laser, and detector, it relies on scanning the delay between the pump pulse and the weak signal field. This necessarily yields long acquisition times, since at any one position of the pump pulse, only a small part of the signal field is sampled.\newline
In this work, we present a method that combines up-conversion and streak cameras, joining the advantages of both techniques, in order to measure quantum dot emission in the infrared. To this aim, we design and demonstrate a sum-frequency generation (SFG) process in Lithium Niobate, which converts an infrared input signal to visible wavelengths while preserving its temporal properties. Notably, this conversion is realised with a cw laser, foregoing the need for ultrashort pump pulses as no scanning is involved.  Further, only certain defined spatial polarization modes are converted, which reduces background compared to direct detection in the IR. \newline
Our work is based on techniques presented in \cite{Allgaier2018} where the information about the temporal shape of the signal was completely lost during the SFG. Here, we design the SFG to precisely conserve the temporal shape of the signal light. \newline
To prove the correct reproduction of the pulse shape, we first convert quantum dot emission at a wavelength of 900~nm where we can detect both the original and the converted light with the streak camera. We show that the conversion process acts as a filter, selecting only the desired spectral band of the quantum dot emission, thus further reducing spurious background photons. Using a 40 mm long crystal, we achieve up to $14 \%$ internal conversion efficiency, which is, in principle, sufficient to improve overall detection efficiencies about two-fold. We then demonstrate a proof-of-concept realisation of a SFG process for higher wavelength regimes, which shows that the method can be flexibly adapted while retaining its favourable noise-filtering properties.

%%%%%%%%%%%%%%%%%%%%%%%%%%%%%%%%%%%%%%%%%%%%%%%%%%%%
%%%%%%%%%%%%%%%%%%%%%%%%%%%%%%%%%%%%%%%%%%%%%%%%%%%%
%%%%%%%%%%%%%%%%%%%%%%%%%%%%%%%%%%%%%%%%%%%%%%%%%%%%
%\section{Sample for sum frequency generation}
\section{Sum-frequency generation for up-conversion detection}

\begin{figure*}[t]
    \centering
    \includegraphics[width=\textwidth]{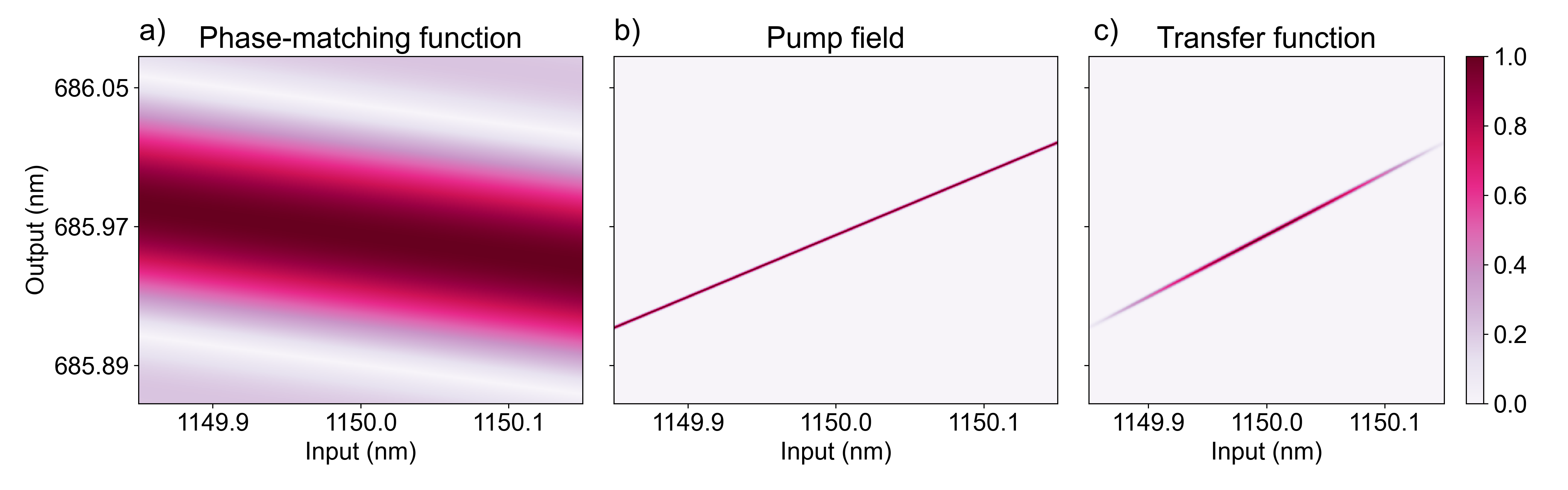}
    \caption{a) Amplitude of the phase-matching function for the engineered type II SFG process. b) Amplitude of the pump field for a continuous-wave laser. c) Amplitude of the transfer function for a process using the phase-matching function in a) and the pump in b). For more information see the text.}
    \label{fig:transfer}
\end{figure*}

    %- If you want to investigate temporal properties, you cannot just use any conversion process
    %- We designed several samples with different properties adapted to the light under investigation\\
    
   %- figure on the sample

SFG is a three wave mixing second order non-linear process.

In this process, a signal and pump field with frequencies $\omega_\text{s}$ and $\omega_\text{p}$ produce a third output field with frequency $\omega_\text{o}=\omega_\text{s} + \omega_\text{p}$.

The spectral properties of the output are given by the transfer function and depend upon the properties of the non-linear material used to produce SFG with the phase-matching function and the pump used for the process. The transfer function can be written as \cite{Eckstein2011}

\begin{equation}
    \Phi(\omega_\text{s},\omega_\text{p}) = \int \text{d}\omega_\text{s} \text{d}\omega_\text{p}\;\alpha(\omega_\text{p})\;\phi(\omega_\text{s} + \omega_\text{p}), \label{eq:transfer_f}
\end{equation}

where $\alpha(\omega_\text{p})$ is the envelope of the pump field and $\phi(\omega_\text{s} + \omega_\text{p})$ the phase-matching function of the process. As depicted in figure \ref{fig:transfer}, the transfer function determines the input-output spectral, and thus, by virtue of the Fourier relationship between time and frequency, also the temporal, characteristics of the SFG. 

In general, a non-engineered transfer function does not ensure the conservation of the signal field spectral characteristics (spectral bandwidth and time duration) upon conversion. 
It is therefore paramount to engineer a transfer function that faithfully reproduces the original signal characteristics at the converted wavelengths. This is realised by a careful adaption of the pump and phase matching properties.
For instance, a 45\textdegree angle of the transfer function, c.f. Fig. \ref{fig:transfer}, produces a 1:1 ratio between signal and output fields and is the ideal condition to achieve a faithful conversion. Note that the spectral acceptance band of the SFG is defined by the phase matching bandwidth and angle in this configuration. Special care must be taken to ensure that the process accepts the complete quantum dot emission, while simultaneously rejecting as much unwanted background as possible. We will demonstrate such a process in the following. In our experiment, the time resolution of the SFG is theoretically infinite due to the continuous-wave pump and is only limited by the streak camera to around 2$\,$ps.\newline
With these considerations in mind, we designed two SFG processes in periodically poled Titanium in-diffused  Lithium Niobate waveguides, matched to the available quantum dots.
The first process transforms quantum dot light at around 900~nm, which is visible to the streak camera, and serves to verify that the temporal and spectral properties are conserved upon conversion.
The second process is designed for quantum dot light that is invisible to the streak camera and demonstrates the flexibility of our approach; its phase-matching and transfer functions are depicted in figure \ref{fig:transfer}. Both processes are engineered to work at around 190\textdegree C to avoid the effects of photo-refraction \cite{Augstov1980}.

%%%%%%%%%%%%%%%%%%%%%%%%%%%%%%%%%%%%%%%%%%%%%%%%%%%%
%%%%%%%%%%%%%%%%%%%%%%%%%%%%%%%%%%%%%%%%%%%%%%%%%%%%
%%%%%%%%%%%%%%%%%%%%%%%%%%%%%%%%%%%%%%%%%%%%%%%%%%%%
\section{Conversion of near-infrared wavelengths}

%%%%%%%%%%%%%%%%%%%%%%%%%%%%%%%%%%%%%%%%%%%%%%%%%%%%
\subsection{Experimental setup}

\begin{figure}[t]
    \centering
    \includegraphics[width=0.43\textwidth]{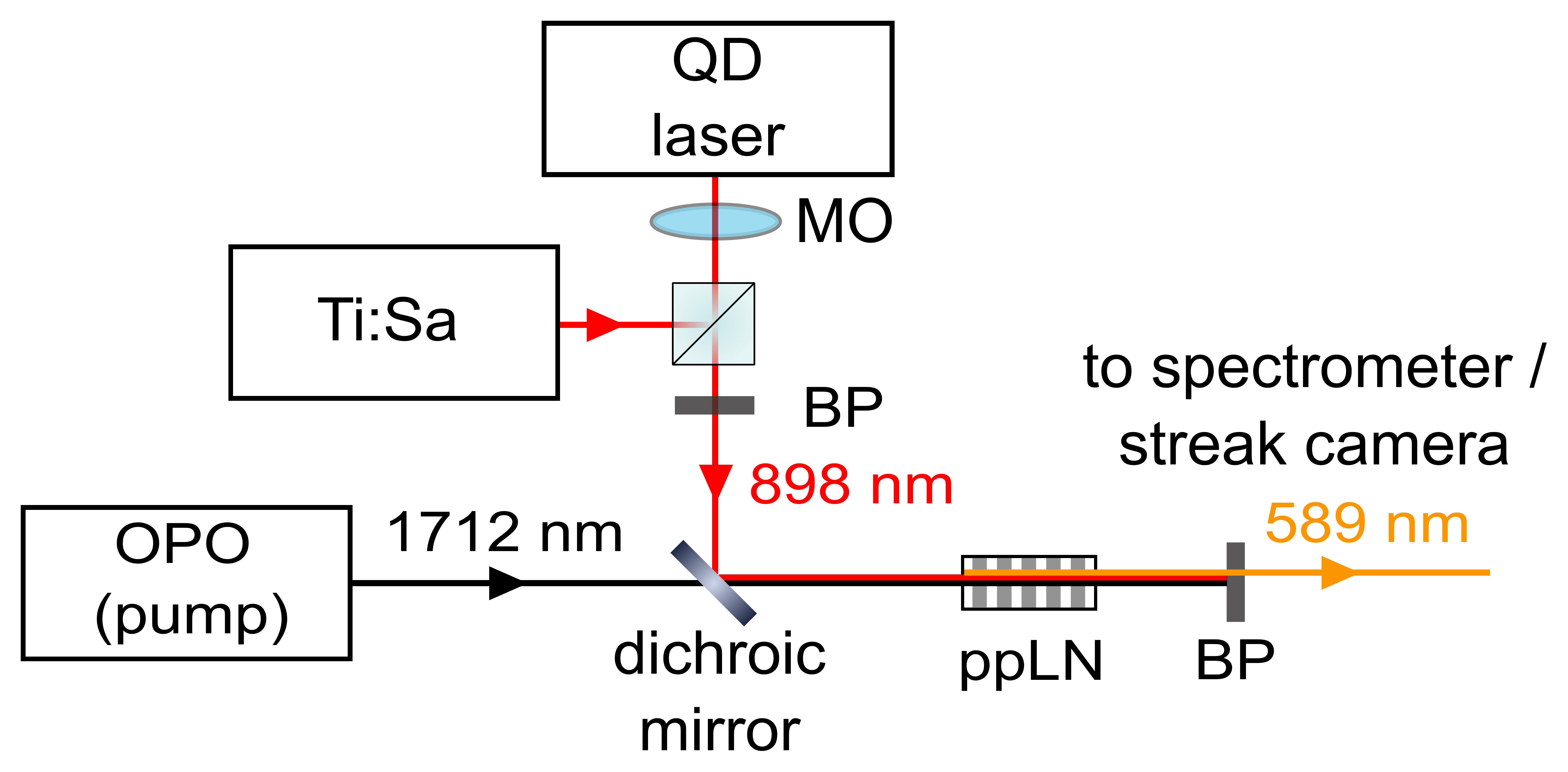}
    \caption{%
        Experimental setup for frequency conversion. Ti:Sa: Ti:Sapphire laser, MO: microscope objective, OPO: optical parametric oscillator, ppLN: periodically poled LiNbO\textsubscript{3} waveguide, BP: bandpass filter. 
    }\label{fig:setup}
\end{figure}

In order to validate the correct replication of the temporal properties, in a first step we converted quantum dot light at a wavelength still accessible for the streak camera. The experimental setup for this task is shown in Fig. \ref{fig:setup}. The continuous-wave pump beam (1712$\,$nm, spectral width $< 1\,\mathrm{MHz}$) was provided by an OPO (Argos 2400). The signal (898$\,$nm) was emitted by Al$_{0.9}$GaIn$_{36}$As quantum dots inside of a GaAs $\lambda$ microcavity, forming a planar microlaser. 26 mirror pairs of GaAs/AlAs create the upper part of the resonator and 33 mirror pairs form the lower one. This microlaser was held in a cryostat at 9$\,$K and excited non-resonantly at 740$\,$nm in a confocal scheme with a 50x microscope objective by a pulsed Ti:Sapphire laser with a repetition rate of 75$\,$MHz. The excitation laser was filtered from the signal with a 900$\,$nm band pass with 10$\,$nm FWHM (Thorlabs FB900-10). After being combined on a dichroic mirror, both the signal and the pump field were focused into the 40$\,$mm long LiNbO\textsubscript{3} waveguide with a poling period of 19\textmu m for the type 0 SFG process. The waveguide was kept at a temperature of 190$^{\circ}$C. After the waveguide, a 600$\,$nm band pass with 10$\,$nm FWHM (Thorlabs FB600-10), which was slightly tilted to adapt to the converted wavelength, removed the pump and any remaining unconverted signal. We detected the SFG up-converted light (589$\,$nm) on a streak camera (Hamamatsu C5680-25) equipped with a S-20 photocathode, having a quantum efficiency of 0.72 \% at 900$\,$nm and 8.16 \% at 590$\,$nm. The device’s deflection circuit was synchronized to the Ti:Sapphire laser repetition rate. For measuring spectra below 1000$\,$nm, we used an Acton SP-2500i spectrometer equipped with a PyLoN:400BR\_eXcelon CCD camera. For wavelengths above 1000$\,$nm, a Stellarnet EPP2000-NIR-InGaAs spectrometer was used. 

%sample: M2754-9-5A from W\"urzburg

%%%%%%%%%%%%%%%%%%%%%%%%%%%%%%%%%%%%%%%%%%%%%%%%%%%%
\subsection{Results}

\begin{figure}[t]
    \centering
    \includegraphics[width=0.5\textwidth]{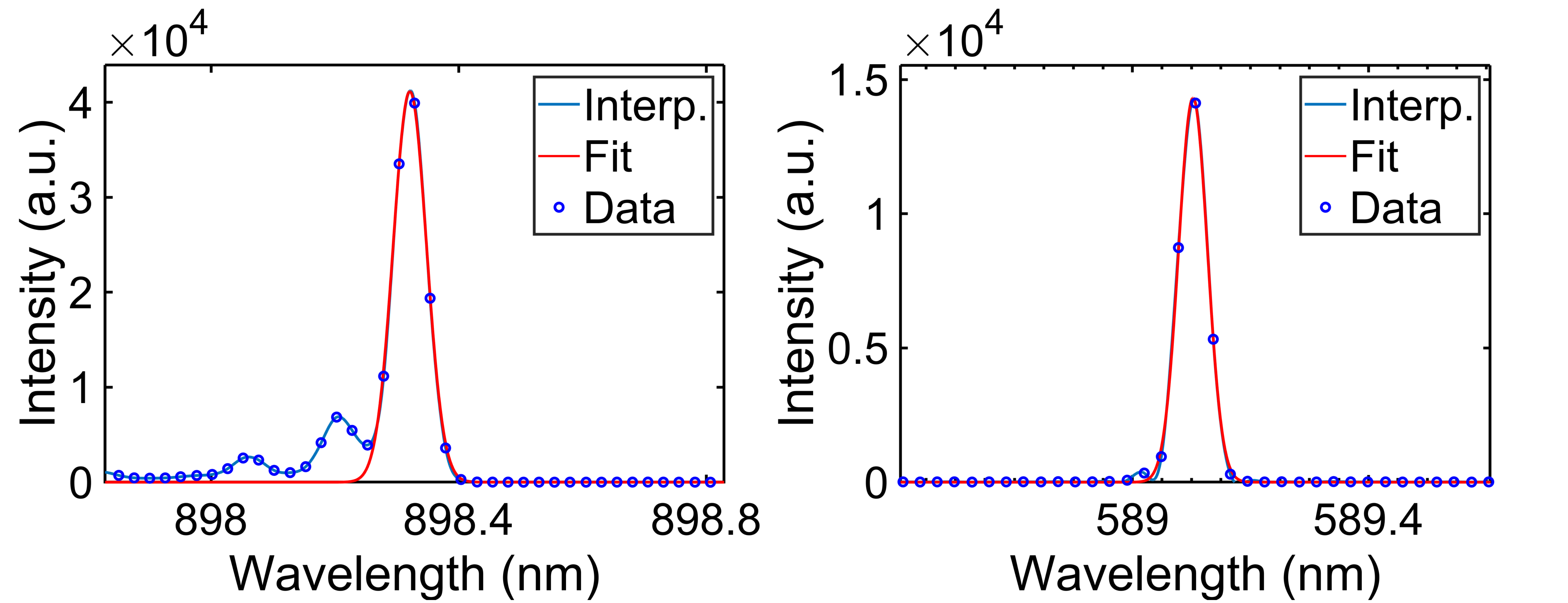}
    \caption{%
        Original (left) and converted (right) spectra of the quantum dot emission.  Blue dots: data, red line: Gaussian fit, blue line: interpolant as guide for the eye. Only the main peak is converted, whereas spurious side peaks are suppressed, showing the selective properties of the SFG process. Note that the intensities are not directly comparable due to different detection sensitivities, filters and losses in the setup.  
    }\label{fig:spectra}
\end{figure}

The original and the converted spectra of the quantum dot laser are compared in Fig. \ref{fig:spectra}. 

The unconverted spectrum consists of several peaks, corresponding to different modes of the microcavity. The most intense peak lies at 898.3$\,$nm with a full width half maximum (FWHM) of 0.06$\,$nm. In contrast to that, the converted spectrum shows only a single peak at 589.1$\,$nm with a FWHM of 0.06$\,$nm. We observe almost no conversion of the less intense peaks. As desired, the SFG acts as a noise filter, selecting only the specific wavelength region for which the phase-matching condition is fulfilled. For this mode, the spectral width is retained during conversion. 

As a figure of merit, we characterized the internal conversion efficiency of the process by measuring the depletion of the quantum dot signal. To this aim, the unconverted signal spectrum was recorded after the waveguide, both with the pump open and blocked. From the spectra, we obtained the peak intensities P\textsubscript{blocked} and P\textsubscript{open}. With these quantities, we calculated the internal conversion efficiency via 

\begin{equation}
\mathrm{Efficiency}= 1 - \frac{P\textsubscript{open}}{P\textsubscript{blocked}}. 
\end{equation}

\begin{figure}[t]
    \centering
    \includegraphics[width=0.45\textwidth]{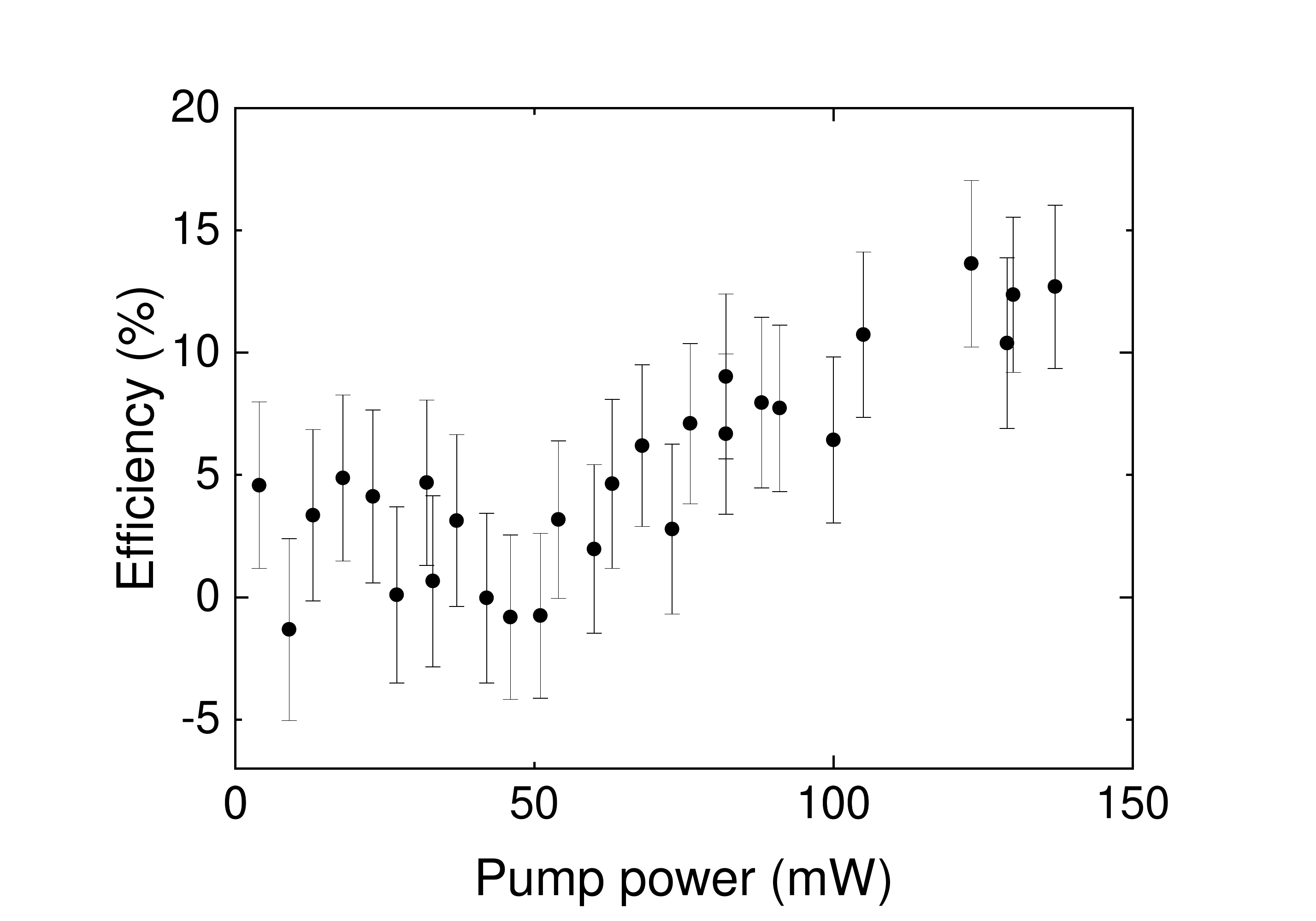}
    \caption{%
        Pump power dependence of the internal conversion efficiency. The pump power was measured after the waveguide. The excitation power of the quantum dots has been kept constant. 
    }\label{fig:efficiency}
\end{figure}

Its dependence upon pump power is shown in Fig. \ref{fig:efficiency}, with the pump power being measured after the waveguide.

According to this, the internal efficiency of the up-conversion depends linearly on pump power in the investigated range of powers and reaches almost up to 14 \%.  This seemingly small number is more than counteracted by the detection efficiencies of the streak camera, which are 0.72 \% at around 900$\,$nm and 8.16 \% at 590$\,$nm, respectively, yielding a potential two-fold increase in detection efficiency when compared to direct detection at the original emission wavelength. Due to additional losses in the setup (e.g., absorption in optical elements, non-perfect waveguide coupling), this was not demonstrated at this point. Instead, care was taken to demonstrate the faithful conversion of the temporal and spectral properties of the signal upon conversion. \newline

\begin{figure*}[t]
    \centering
    \includegraphics[width=0.8\textwidth]{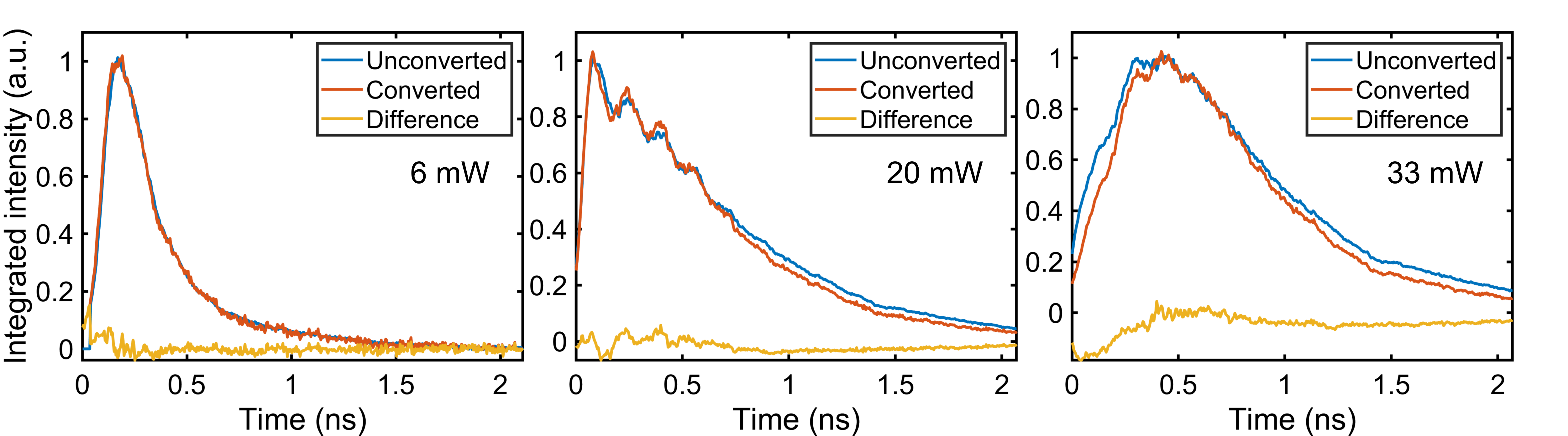}
    \caption{%
        Temporal wave forms of the original (blue) and converted (red) quantum dot emission for different excitation powers. The curves are normalized and shifted so that their maxima coincide. Yellow: Intensity difference. Obviously, the frequency conversion process accurately retains the temporal properties of the input light. 
    }\label{fig:pulses}
\end{figure*}

The key results of our proof-of-principle experiment are the temporal pulse shapes of the original and the up-converted quantum dot emission, measured with the streak camera. They are compared in Fig. \ref{fig:pulses}. In the unconverted case, all the spectral components of the signal shown in Fig. \ref{fig:spectra} (left) are sent to the streak camera, bypassing the waveguide.\newline 
The pulse shape of the quantum dot emission varies with excitation power. We find that the up-converted pulse shape reproduces the original one accurately. Thus, the temporal properties of the quantum dot emission are preserved during the conversion, even while spurious side peaks in the spectral domain are not converted. This leads to the conclusion that the SFG process coherently filters noise from the input.\newline
Only for higher excitation powers, the converted pulse is slightly shorter than the unconverted one, while still preserving the shape qualitatively. This might be caused by the spectral components of the signal that are not converted and are not removed by the 900$\,$nm-band pass, as they lie too closely to the main peak. For higher excitation powers, these might also be detected by the streak camera and might arrive later than the emission from the main peak.  \newline  

%%%%%%%%%%%%%%%%%%%%%%%%%%%%%%%%%%%%%%%%%%%%%%%%%%%%
%%%%%%%%%%%%%%%%%%%%%%%%%%%%%%%%%%%%%%%%%%%%%%%%%%%%
%%%%%%%%%%%%%%%%%%%%%%%%%%%%%%%%%%%%%%%%%%%%%%%%%%%%
\section{Conversion of 1145 nm light}

\begin{figure}[t]
    \centering
    \includegraphics[width=0.50\textwidth]{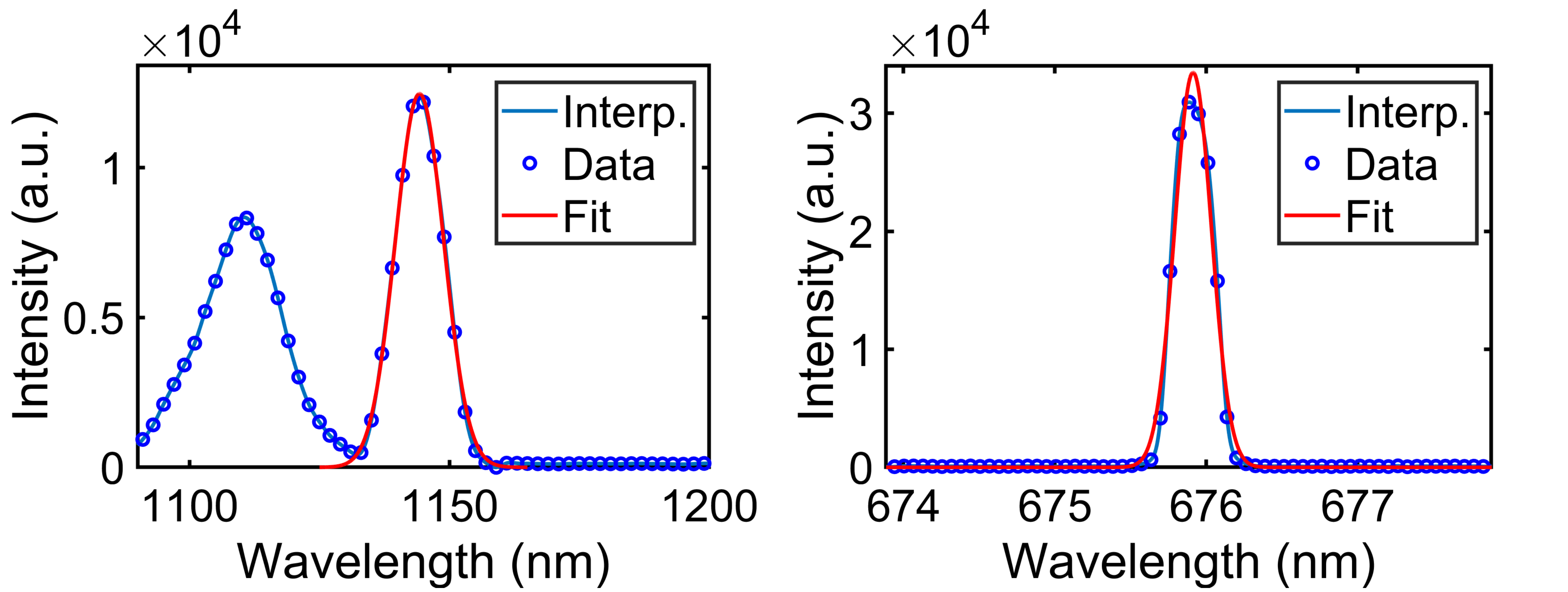}
    \caption{%
        Original (left) and converted (right) spectra of the 1145$\,$nm laser light. Blue dots: data; red line: Gaussian fit; blue line: interpolant as guide for the eye. Note that the intensities are not directly comparable due to different spectrometers and losses in the setup.  
    }\label{fig:spectra-new}
\end{figure}

While the previous results validate the capability of our up-conversion process by converting wavelengths where streak cameras still possess small, but non-zero efficiency, the final goal is to convert wavelengths that are invisible for streak cameras. To achieve this, we designed an optimized type II SFG process with a poling period of 7\textmu m, which combines light at 1150$\,$nm with a pump at around 1700$\,$nm to generate converted light at around 686$\,$nm. To prove the applicability of this process, we converted test light from a continuous-wave diode laser (Toptica TA Pro) at 1145$\,$nm. The setup was similar as the one shown in Fig. \ref{fig:setup}, but the pump was now set to 1649.2$\,$nm to fulfil the phase-matching condition. The resulting spectra are shown in Fig. \ref{fig:spectra-new}. 

The unconverted signal consists of a broad emission around 1110$\,$nm, produced by the ASE of the diode laser, and a peak of coherent emission at 1145$\,$nm. Its broad FWHM of 11$\,$nm is determined by the resolution of the EPP2000-NIR-InGaAs spectrometer, whereas the original laser linewidth lies in the MHz range and can not be resolved. The converted signal has a peak at 675.9$\,$nm with a FWHM of 0.3$\,$nm. We observe again that the central peak is converted to the desired output wavelength.  Furthermore, as before, side peaks are suppressed, demonstrating the selective nature of the conversion.\newline
Let us briefly discuss the effect of conversion at this wavelength for streak cameras. At the converted wavelength 676~nm, an S-25 photocathode has ca 30~mA/W radiant sensitivity \cite{Hamamatsu2018}. In contrast to that, at 1145~nm, S1 photocathodes, which are better for the infrared, have a radiant sensitivity of ca 0.1~mA/W \cite{Hamamatsu2018}, while InP/InGaAs photocathodes have a radiant sensitivity of ca 1~mA/W at 1145~nm \cite{Hamamatsu2015}). Thus, when comparing a measurement of the converted signal with an S-25 to a measurement of the original signal with an S-1 photocathode, we gain a factor of ca 300; when comparing to a InP/InGaAs photocathode, the factor is still 30. Assuming an internal conversion efficiency of 14 \% as demonstrated in the proof-of-concept experiment, the system can accept up to 95 \% additional loss in the first case and up to 52 \% additional loss in the second case to still outperform direct streak camera detection twofold at this wavelength. We note that this huge loss resilience emphasizes the strength of combining SFG with streak camera detection for near- to mid-infrared wavelengths: even for non-optimized systems, as would be the case in real-world applications - one finds a significant performance enhancement while retaining all the upsides of streak camera detection for temporal signal analysis.

%%%%%%%%%%%%%%%%%%%%%%%%%%%%%%%%%%%%%%%%%%%%%%%%%%%%
%%%%%%%%%%%%%%%%%%%%%%%%%%%%%%%%%%%%%%%%%%%%%%%%%%%%
%%%%%%%%%%%%%%%%%%%%%%%%%%%%%%%%%%%%%%%%%%%%%%%%%%%%
\section{Conclusion}\label{sec:Conclusion}

We have demonstrated that careful design of frequency conversion processes can fundamentally enhance quantum dot technologies. In particular, we have designed and experimentally confirmed two SFG processes that preserve the temporal properties of quantum dot emission while converting it to the visible range, where streak-cameras can be used at higher efficiencies. Overall, we demonstrate our system to achieve up to a twofold increase in the detection efficiency of quantum dots at 900$\,$nm and infinite enhancement for emission at 1150$\,$nm while maintaining spectral properties. The system is resilient to losses and already outperforms S-1 and InP/InGaAs photocathodes for telecom signal detection. Furthermore, the SFG process acts as a spectral and polarization filter for the quantum dot emission, thanks to the spatial single-mode propagation of the waveguides, the type II conversion process between orthogonal polarisation and the transfer function. Finally, we would like to emphasize the major strength of our novel SFG technique, that is directly reproducing the sought-after pulse shape without any need to deconvolute the signal or scan a delay line. This technique presents a promising starting point for further applications in the  spectroscopy of infrared lightsources.

%%%%%%%%%%%%%%%%%%%%%%%%%%%%%%%%%%%%%%%%%%%%%%%%%%%%
%%%%%%%%%%%%%%%%%%%%%%%%%%%%%%%%%%%%%%%%%%%%%%%%%%%%
%%%%%%%%%%%%%%%%%%%%%%%%%%%%%%%%%%%%%%%%%%%%%%%%%%%%
\begin{acknowledgements}
    The authors would like to thank Christian Schneider and Sven H\"ofling for providing the microlaser sample. We gratefully acknowledge funding through the Deutsche Forschungsgemeinschaft (DFG, German Research Foundation) via the transregional collaborative research center TRR~142 (project C01, grant no. 231447078).  
\end{acknowledgements}

%%%%%%%%%%%%%%%%%%%%%%%%%%%%%%%%%%%%%%%%%%%%%%%%%%%%
%%%%%%%%%%%%%%%%%%%%%%%%%%%%%%%%%%%%%%%%%%%%%%%%%%%%
%%%%%%%%%%%%%%%%%%%%%%%%%%%%%%%%%%%%%%%%%%%%%%%%%%%%
\appendix

%%%%%%%%%%%%%%%%%%%%%%%%%%%%%%%%%%%%%%%%%%%%%%%%%%%%

%%%%%%%%%%%%%%%%%%%%%%%%%%%%%%%%%%%%%%%%%%%%%%%%%%%%
%%%%%%%%%%%%%%%%%%%%%%%%%%%%%%%%%%%%%%%%%%%%%%%%%%%%
%%%%%%%%%%%%%%%%%%%%%%%%%%%%%%%%%%%%%%%%%%%%%%%%%%%%

\end{document}